\PassOptionsToPackage{hyphens, spaces}{url}

\documentclass[acmsmall]{acmart}

\usepackage{multirow}
\usepackage[hyphens,spaces]{url}
\usepackage{caption}
\usepackage{subcaption}
\usepackage{tabularx}
\usepackage{makecell}

\AtBeginDocument{%
  \providecommand\BibTeX{{%
    \normalfont B\kern-0.5em{\scshape i\kern-0.25em b}\kern-0.8em\TeX}}}

\author{Rijul Magu}
\affiliation{
  \institution{Georgia Institute of Technology}
  \city{Atlanta}
  \state{Georgia}
  \country{USA}
  }
\email{rijul.magu@gatech.edu}

\author{Nivedhitha Mathan Kumar}
\affiliation{
  \institution{Georgia Institute of Technology}
  \city{Atlanta}
  \state{Georgia}
  \country{USA}
  }
\email{nive.mathan21@gmail.com}

\author{Yihe Liu}
\affiliation{
  \institution{University of Washington}
  \city{Seattle}
  \state{Washington}
  \country{USA}
  }
\email{yliu297@cs.washington.edu}

\author{Xander Koo}
\affiliation{
  \institution{Georgia Institute of Technology}
  \city{Atlanta}
  \state{Georgia}
  \country{USA}
  }
\email{xander@gatech.edu}

\author{Diyi Yang}
\affiliation{
  \institution{Stanford University}
  \city{Stanford}
  \state{California}
  \country{USA}
  }
\email{diyiy@stanford.edu }

\author{Amy Bruckman}
\affiliation{
  \institution{Georgia Institute of Technology}
  \city{Atlanta}
  \state{Georgia}
  \country{USA}
  }
\email{asb@cc.gatech.edu}

\renewcommand{\shortauthors}{Rijul Magu et al.}

\begin{document}

\setcopyright{acmlicensed}
\acmJournal{PACMHCI}
\acmYear{2024} \acmVolume{8} \acmNumber{CSCW2} \acmArticle{404} \acmMonth{11} \acmPrice{15.00}\acmDOI{10.1145/3686943}

\title{Understanding Online Discussion Across Difference: Insights from Gun Discourse on Reddit
}

\begin{abstract}
    When discussing difficult topics online, is it common to meaningfully engage with people from diverse perspectives? Why or why not? Could features of the online environment be redesigned to encourage civil conversation across difference? To investigate these questions, we need to explore them in a particular context. In this paper, we study discussions of gun policy on Reddit, with the overarching goal of developing insights into the potential of the internet to support understanding across difference. We use two methods: a clustering analysis of Reddit posts to contribute insights about \emph{what} people discuss, and an interview study of twenty Reddit users to help us understand \emph{why} certain kinds of conversation take place and others don't.
    We find that the discussion of gun politics falls into three groups: conservative pro-gun, liberal pro-gun, and liberal anti-gun. Each type of group has its own characteristic topics. While our subjects state that they would be willing to engage with others across the ideological divide, in practice they rarely do.
    Subjects are siloed into like-minded subreddits through a two-pronged effect, where they are simultaneously pushed away from opposing-view communities while actively seeking belonging in like-minded ones.
    Another contributing factor is Reddit's ``karma'' mechanism: fear of being downvoted and losing karma points and social approval of peers causes our subjects to hesitate to say anything in conflict with group norms. The pseudonymous nature of discussion on Reddit plays a complex role, with some subjects finding it freeing and others fearing reprisal from others not bound by face-to-face norms of politeness. Our subjects believe that content moderation can help ameliorate these issues; however, our findings suggest that moderators need different tools to do so effectively. We conclude by suggesting platform design changes that might increase discussion across difference.

\end{abstract}

\ccsdesc[500]{Human-centered computing~Empirical studies in collaborative and social computing}

\keywords{social media, gun politics}

\maketitle

\section{Introduction}
Is it common for people on social media to have discussions about controversial issues with others they don't already agree with? When those conversations occur, do they often lead to greater mutual understanding?
Could deepening our understanding of these phenomena help us to design strategies for ameliorating affective polarization?
Affective polarization refers to the animosity felt by partisans toward their perceived political opponents \cite{iyengar2012affective}, a phenomenon that is only further fueled by the lack of direct social contact \cite{bail2022socialMediaPrism}. But affective polarization does not necessarily correspond with an actual difference in policy positions. Studies show that people tend to overestimate the extent to which their beliefs differ from other people \cite{baladassarri2008partisans, dimaggio1996socialAttitudes, lelkes2016massPolarization} by about 20\% \cite{levendusky2015misperceptions}. Thus, people who ostensibly disagree on controversial subjects could potentially find areas of common ground if they had the opportunity to speak respectfully with one another.
In order to foster understanding across difference and reduce affective polarization online, it is important to understand how to facilitate cross-partisan social contact through civil discussion of controversial issues.
What factors support and impede cross-partisan civil discussion online? In this research, we pick a controversial topic, gun politics, and explore how people engage with this topic online. 

To study discussion across political difference, we must examine it in a particular context. We selected gun politics as the focus of our study because prior work has found it to be one of the most polarizing contemporary political issues \cite{wojatzki2018quantifying}, and sizable communities dedicated to discussion of gun politics exist across multiple sides of the political divide on Reddit, the platform of choice for this study. Gun politics is our \textit{place} of study, but not our \textit{object} of study. Anthropologist Clifford Geertz wrote that social scientists do not study villages, they study \emph{in} villages \cite{geertz}. In the village of gun politics, our goal is to study to what extent people have meaningful conversations with others they disagree with, and why this is so.
Our study is guided by four research questions:

\begin{quote}
    \textbf{RQ1:} \textit{What are common themes of online gun politics discussions on Reddit?}
\end{quote}

\begin{quote}
    \textbf{RQ2:} \textit{What motivates individuals to contribute to contentious discussions about guns online? }
\end{quote}

\begin{quote}
    \textbf{RQ3:} \textit{What are the barriers to meaningful political discourse surrounding the topic of guns?}
\end{quote}

\begin{quote}
    \textbf{RQ4:} \textit{What are the  advantages and disadvantages of Reddit as a platform for political discourse? }
\end{quote}

At the time of this writing, Reddit ranks among the world's most visited social media websites \cite{SimilarWebSocialMedia}.
The platform serves as a ``Third Place'' \cite{oldenburg1982third} for a multitude of interest-based online communities ranging from sports to humor to politics.  
However, Reddit suffers from an echo-chamber problem on issues pertaining to politics \cite{bond2022political}. Users rarely engage with those they disagree with.
Prior work has shown that while participants who contribute to \textit{homogeneous} subreddits (i.e., those with specific political leanings, such as r/hillaryclinton and the now-defunct r/The\_Donald) do interact on \textit{cross-cutting} subreddits as well, they rarely engage with their partisan counterparts on the same threads \cite{an2019political}. This motivates a need to understand the dynamics that hinder controversial political interactions between groups, even in spaces that encourage diverse participation. 

Our work aims to understand gun politics communication through two complementary angles: (1) by exploring themes of discussion expressed through content titles of Reddit posts and (2) by analyzing the self-reported experiences of Reddit contributors. We first conducted a clustering-based experiment to learn about general themes and patterns of community discussions across gun-related subreddits as expressed through post titles. However, posted content does not fully explain users' motives behind  contributions and the barriers they face when attempting to interact with others. Therefore, for our second analysis, we carried out an interview study to learn about the experiences of participants who contribute to such subreddits. 
Based on our findings, we provide design insights into what sociotechnical features might better support a productive discussion of controversial topics on Reddit and other platforms. 

Our research indicates that gun policy subreddits fall into three main categories: conservative pro-gun, liberal pro-gun, and liberal anti-gun (see Section \ref{content-analysis}). There are no conservative anti-gun groups, as far as we know. In this study, we analyze the topics discussed on eight subreddits and present our insights from interviews with users who posted in these communities. 
We focus on groups where people primarily discuss gun policy, not just gun hardware. Because gun politics differ regionally, we have chosen to focus on the US context. 

Our results reveal that communities on opposing sides of the gun debate engage in quite different discussions.
On one hand, while conservative pro-gun communities tend to focus on themes surrounding electoral politics and court rulings relating to gun control, liberal anti-gun communities tend to hold conversations around empirical studies of gun violence and news events. 

In addition, our findings indicate that there exists a two-pronged, push-pull effect that pushes our subjects away from opposing-view communities, due to perceived hostility or incompatible values, while also pulling them toward like-minded gun politics subreddits as they seek belonging among people with similar experiences.
Although cross-ideology interactions are currently limited, users state that they would be willing to communicate online with those whom they disagree with, provided the existence of a platform with certain features and ideology-agnostic rules to promote civil interactions.

From a platform design standpoint, we observe that site design choices on Reddit related to downvoting, brigading, anonymity, and moderation play a crucial role in enabling discussions between groups with differing views and social norms. For instance, concerns about possible loss of karma (Reddit's measure of user reputation) or soft removal of content (in the form of auto-collapsing threads) as a result of mass downvoting deter users from contributing to opposing-ideology communities. Our findings point to sociotechnical features that might better support cross-ideology communication.
Barriers to cross-ideology communication are not necessarily caused by the platform's design, nor can they be solved by it; however, deliberate design choices may nonetheless have the potential to support more harmonious conversation across ideological difference.

\section{Related Work}
We set the stage with a review of research on the discussion of controversial issues online. Next, we provide an overview of the issue of guns in America. Finally, we introduce what is known about the discussion of guns and gun policy on the internet.

\subsection{Discourse about Controversial Issues}

Cross-ideology discourse about controversial issues may play a role in reducing polarization. Since affective polarization is driven by lack of direct contact with perceived political opponents \cite{bail2022socialMediaPrism} and people tend to overestimate how different their beliefs are from those of other people \cite{baladassarri2008partisans, dimaggio1996socialAttitudes, lelkes2016massPolarization}, there may be unexplored points of agreement among partisans that would be revealed through civil discourse. Discussion of controversial issues is also important for democracy. For instance, participation in controversial discussions among people with differing viewpoints is indicative of increased political participation \cite{mcleod2001values}. Discussion helps citizens to understand and tolerate the views of others \cite{hess2002discussing} and arrive at carefully considered opinions \cite{price2002does}. 

A commonly advocated method for achieving democratic discourse is to promote intergroup interactions. Exposure to dissenting views is found to increase political tolerance \cite{mutz2002cross}, increase political participation \cite{min2018all}, and improve the ability to understand others' views \cite{price2002does}. But, given that people may stay away from intergroup contact to avoid cognitive dissonance \cite{festinger1962theory} and have biases as a result of affective polarization \cite{iyengar2012affective}, researchers have identified strategies that facilitate efficient cross-cutting exposure. Clearing misperceptions regarding outgroup (communities with views perceived as antithetical to one's own) stereotypes makes people perceive outgroup members as less extreme and reduces how socially distant they feel toward the outgroup. \cite{ahler2018parties}. Additionally, priming a common identity (for example, American) over partisan identity can increase positivity toward the outgroup \cite{levendusky2018americans}. Likewise, ``minimizing the salience of partisanship in structured bipartisan networks'' by removing cues such as party logos helped reduce bipartisan bias and gain a better understanding of contentious data \cite{guilbeault2018social}.

However, there are obstacles to controversial discussions. Prior work suggests that higher amounts of controversy reduce the likelihood of conversation \cite{chen2013and}. Discussions of controversial issues can provoke an emotional reaction in participants \cite{hess2002discussing}, and people often find disagreement uncomfortable \cite{matz2005cognitive}.  
Furthermore, studies indicate that it is more discomforting when participants have to arrive at a consensus than merely share their opinions \cite{matz2005cognitive}, which further raises the difficulty of having meaningful dialogue across ideological boundaries. 

Civility and respect among interlocutors are important for cultivating cross-ideology discourse.
A successful dialogue includes the ability of participants to ``provide fully developed arguments for their own position and take seriously and respond to the arguments of others'' \cite{ferree2002shaping}. On the other hand, incivility in outgroup interactions can reduce one's willingness to engage \cite{hwang2008does} and increase the likelihood of provoking an uncivil response \cite{masullo2017online}.

Social media has complex impacts on discourse. On one hand, it reduces discomfort by offering anonymity. Engagement increases when requirements for self-disclosure are reduced \cite{chen2013and}. Social media also supports reflective discourse \cite{coleman2009internet} that can help develop a deeper understanding of issues in comparison to simple reception of factual information \cite{mcleod2001values}. On the other hand, anonymity online can reduce civility \cite{santana2014virtuous,rowe2015civility}. 

\subsection{The Gun Debate in the US}

In the United States, regulation of guns is a hotly contested issue \cite{blanco2016gun, lacombe2019gun}. 
While pro-gun supporters often favor individualistic or hierarchical worldviews (by focusing on individual self-sufficiency), those on the anti-gun end of the spectrum typically hold egalitarian or communitarian worldviews (by focusing on public safety) \cite{braman2006overcoming}. These cultural worldviews are also reflected in the groups’ respective interpretations of the Second Amendment (2A) to the US Constitution. The exact wording of the second amendment to the US Constitution says, ``A well regulated Militia being necessary to the security of a free State, the right of the people to keep and bear Arms shall not be infringed'' \cite{2ndAmendment}. While pro-gun supporters consider gun control as ``an infringement of individual rights'' provided by the 2A,  supporters of increased gun regulation typically believe that the preceding clause about ``a well regulated militia'' means that those rights are limited. \cite{celinska2007individualism}. 

Scholars also attempt to explain the division between pro- and anti-gun supporters using  
connotations or associations related to guns. Among gun advocates, guns act as symbols of cultural heritage \cite{metzl2019guns}, freedom \cite{warner2021guns}, and American identity \cite{jouet2019guns}. On the other side, among critics, guns symbolize social hierarchies, an indifference to the well-being of others \cite{braman2006overcoming}, and “markers of collective trauma” \cite{metzl2019guns}. Past work has studied the association of guns with masculinity. 
For instance, \citet{warner2022provide} found that stereotypical masculine ideals 
were associated with protective ownership of guns.
 
One way to explain the divisiveness of gun issues is through the use of Social Identity Theory.
According to the theory, people perceive themselves as belonging to a group based on its prototypical characteristics \cite{ashforth1989social}. These characteristics may be used by group members to define their personal identities and prompt them to perform activities that are congruent with the group identity.  
\citet{lacombe2019gun} identified that factors such as participation in gun-related activities, contact with the NRA, and the possession of a sense of threat to gun rights are linked to a higher sense of identification as a gun owner. 
In other words, for many gun owners, the weapons they own are not just useful tools for self-protection or sport, but an expression of who they are.
Furthermore, people carrying negative stereotypes of gun owners have higher chances of supporting bans \cite{bankston1990influence}, which aligns with observations by \citet{ashforth1989social} which suggest that an increased salience of in-group identification is related to increased stereotyping of the out-group.

Despite the divisive nature of the issue, people with different ideological positions generally agree on certain policy solutions. According to the survey by Pew Research Center in 2021, there is bipartisan support for regulations including the prevention of gun purchases by those with mental illnesses and the expansion of background checks \cite{doherty2021amid}. A survey in 2015 by \citet{barry2015two} also concluded that a majority of the respondents supported background checks. However, such ``areas of agreement are seemingly ignored or downplayed'' on related online resources which are hypothesized to, in turn, increase polarization \cite{fisher2018social}.

Gun ownership is a major predictor of one's position on gun control \cite{kleck2009people, celinska2007individualism, wozniak2017public}.Studies show that gun owners are three times more likely to oppose gun control measures \cite{celinska2007individualism}. Other key predictors include race, gender, and political affiliation, with white, male, and conservative affiliations being most associated with higher gun ownership and lower preference for gun control measures \cite{oraka2019cross, wozniak2017public, gresham2020owns}. 
Factors such as fear of crime \cite{jiobu2001lack, warner2020matter,bankston1990influence} and victimization \cite{warner2020matter,gresham2020owns,bankston1990influence}, on the other hand, gathered mixed results on whether they play a significant role in prediction.
While many of these predictors continue to be valid, the demographics appear to be changing among the new gun owners. A recent study found that among those who purchased a gun in 2019--2021, approximately 48\% were women --- 20\% higher than the share of women among existing gun owners \cite{miller2022firearm}. The researchers also observed a lower number of non-Hispanic White purchasers among the new gun owners (55\% vs. 74\% among existing owners). A similar study found that new gun owners of 2015--2020 had higher odds of being liberals when compared to those who already owned a gun \cite{wertz2018differences}. In addition, they were more likely to purchase a gun for the purpose of protection. The finding is supported by Pew Research Center's surveys that collected the reasons for gun ownership. On comparing the 1999 and 2013 survey results, `protection' (22\% increase) surpassed `hunting' (17\% drop) as the top cited reason \cite{dimock1999own}.

Social media and internet usage is tied strongly to beliefs about democratic values and offline political participation \cite{swigger2013online, zhuravskaya2020political}. 
Given that social gun culture could influence ownership \cite{kalesan2016gun}, online engagements within gun-related echo chambers may contribute to evolving positions around gun culture.

Online discussions of gun politics are polarized. \citet{demszky2019analyzing} found that Twitter discourse around mass shooting events tends to be polarized across dimensions such as framing and topic choice.
The study identified that the differential use of framing devices (for instance the term ``terrorist'') by Republicans and Democrats drives the gap. A similar study by \citet{barbera2015tweeting} on the Newtown shooting observed that polarization increased on Twitter with time as the focus of discussion shifted ``from the tragedy to a debate over gun-control policy.'' A qualitative analysis of online information on gun rights and gun control by \citet{fisher2018social} theorized that the simplification of arguments to a state where no common ground is acknowledged lent to the polarization of the issue.

\subsection{Political Discourse on Reddit}
 Reddit differentiates itself from other major social network platforms due to a combination of affordances suited to community building. It has a public message-board interface partitioned into individual communities or \textit{subreddits}. Participation within subreddits is pseudonymous --- while context collapse can indirectly reveal portions of user identity \cite{triggs2021context}, specific public user details are not publicly linked. Prior work suggests anonymity on Reddit and the associated disinhibition effect can encourage self-disclosure and support \cite{de2014mental, ammari2019self}. Each subreddit is moderated by its own moderation team. Although participation is subject to global Reddit rules, each subreddit commonly enforces additional rules that align with community norms. Research indicates that effective distributed moderation systems of this nature can help support community participation \cite{lampe2014crowdsourcing}. For ranking content, Reddit incorporates an upvote-downvote system. Each post or comment can be upvoted or downvoted by participants and higher rated content is bubbled to the top of the page. At the same time, content that is heavily downvoted can be flagged and collapsed.

Importantly, Reddit acts as a suitable space for interactions \cite{soliman2019characterization} across a spectrum of political leanings.  Community themes range from having broad discussion mandates (\textit{r/politics}) and representations of major political ideologies (\textit{r/Conservative}, \textit{r/Liberal}) to more niche issue-specific settings (\textit{r/abortion}, \textit{r/liberalgunowners}). One form of categorization of such communities compares \textit{homogeneous} (where participants belong to relatively confirmed interest groups such as \textit{r/hillaryclinton} and \textit{r/The\_Donald})  against \textit{cross-cutting} subreddits (communities with diverse, leaning agnostic participation such as \textit{r/politics} and \textit{r/news}) \cite{an2019political}. While participants across Reddit usually exhibit participation in only one community \cite{buntain2014identifying} prior work suggests that participants of politically opposed homogeneous communities do contribute to cross-cutting subreddits --- however, few directly engage with one another on the same threads \cite{an2019political}. Interestingly, Kaw et al. also show that when users do engage in cross-cutting communities, they shift communication strategies (for instance Clinton supporters use more persuasive language when interacting with Trump supporters \cite{an2019political}. In addition, political discourse invites toxicity, with political comments on average being more offensive than non-political ones \cite{nithyanand2017measuring}. A combination of these factors, therefore, inhibits positive cross-ideology conversations on controversial political topics on Reddit. Furthermore, prior work indicates that contentious interactions on Reddit can be predicted using a combination of linguistic and account-level markers \cite{beel2022linguistic, hessel2019something, guimaraes2019analyzing}  
Thus, to gain a holistic understanding of online discussions of gun politics, studies involving analysis of content and user motives behind posting are needed.

\section{Methods}
To understand the current civic discussion of gun-related issues online, we used two modes of analysis: (1) a quantitative content-clustering approach that analyzes gun politics subreddits and (2) an interview study with users of such subreddits.

Our choice of methods is driven by a need to holistically examine the varied landscape of gun communication online. Our quantitative approach lets us investigate \emph{what} people are discussing. Clustering is an established method to uncover underlying discussion topics within online communities \cite{park2018examining, mitra2015credbank}. The approach enables the study of posting patterns and the common themes that emerge across a large number of posts.

To go beyond what people are discussing to explore \emph{why} particular patterns of participation emerge, we conducted a qualitative in-depth investigation. By interviewing members of these gun subreddits, we can learn about participants' lived-experiences, motivations and challenges. Semi-structured interviews have established validity in online community research as evidenced by prior work \cite{bryant2005becoming, triggs2021context}.

\subsection{Content Analysis of Gun Subreddits} \label{content-analysis}

In the first analysis, we assessed the content of posts on popular subreddits centered on gun discussions to help answer \textbf{RQ1}. To build our dataset, we first queried the Reddit API to search for “gun politics” subreddits. The retrieval was limited to the top 100 results and was based on the title and subreddit descriptions. Next, the first author coded the resultant subreddits based on filter criteria described below. A second round of coding was conducted independently by the a co-author following the same guidelines, with 100\% agreement.  
The search yielded a list which included (1) irrelevant or tangentially related communities that occasionally discussed gun content (e.g., \textit{r/florida} and \textit{r/politics}) and (2) very small communities (fewer than 500 members). Upon filtering these out, we were left with 48 subreddits that directly discussed guns.

We conducted finer categorization of the remaining subreddits on the basis of support for political discussion. Several communities explicitly enforce variations of a \textit{“No Politics”} rule (e.g., \textit{r/ak47)}. Some also severely restrict the amount of political communication (e.g., \textit{r/guns} allows discussion only on specified bi-weekly politics threads). Others do not explicitly prohibit or encourage gun politics content but have negligible, if any, posts that relate to this theme. These are typically subreddits that have extremely narrow mandates (e.g., \textit{r/300BLK} and \textit{r/3mm}). We categorize these subreddits as \textit{Non-Political}.
A number of gun subreddits pertain to topics restricted to geographical locations. These niche subreddits are limited in scope and do not extend discussion to general US politics (e.g., \textit{r/TexasGuns}). In some cases, the subreddits are based outside of the United States (e.g., \textit{r/canadaguns}). We categorize all such subreddits as \textit{Region-Specific}.

Finally, a category of subreddits can be categorized as \textit{Political Subreddits}. These consist of 10 communities that explicitly encourage political discussion. From this set, we selected the following 8 subreddits to arrive at our final analysis set: \textit{r/gunpolitics}, \textit{r/progun}, \textit{r/progunyouth}, \textit{r/shitguncontrollerssay}, \textit{r/guncontrol}, \textit{r/GunsAreCool}, \textit{r/liberalgunowners} and \textit{r/2Aliberals}. At the time of writing, \textit{r/shitguncontrollerssay} has been banned for violating Reddit's Moderator Code of Conduct.

We note that two political subreddits were excluded as they were found to not be suitable for the study. The first, \textit{r/dgu} (a subreddit on defensive gun use), supports political discussion, however, posts are primarily required to only be links to news articles and not discussions. The second, \textit{r/SocialistRA}, is primarily intended to be a discussion space for members of a specific organization --- namely the Socialist Rifle Association. As such, its mandate is different from the other political subreddits that are not subject to external organizational considerations.

\begin{figure}[h]
  \centering
  \includegraphics[width=0.8\linewidth]{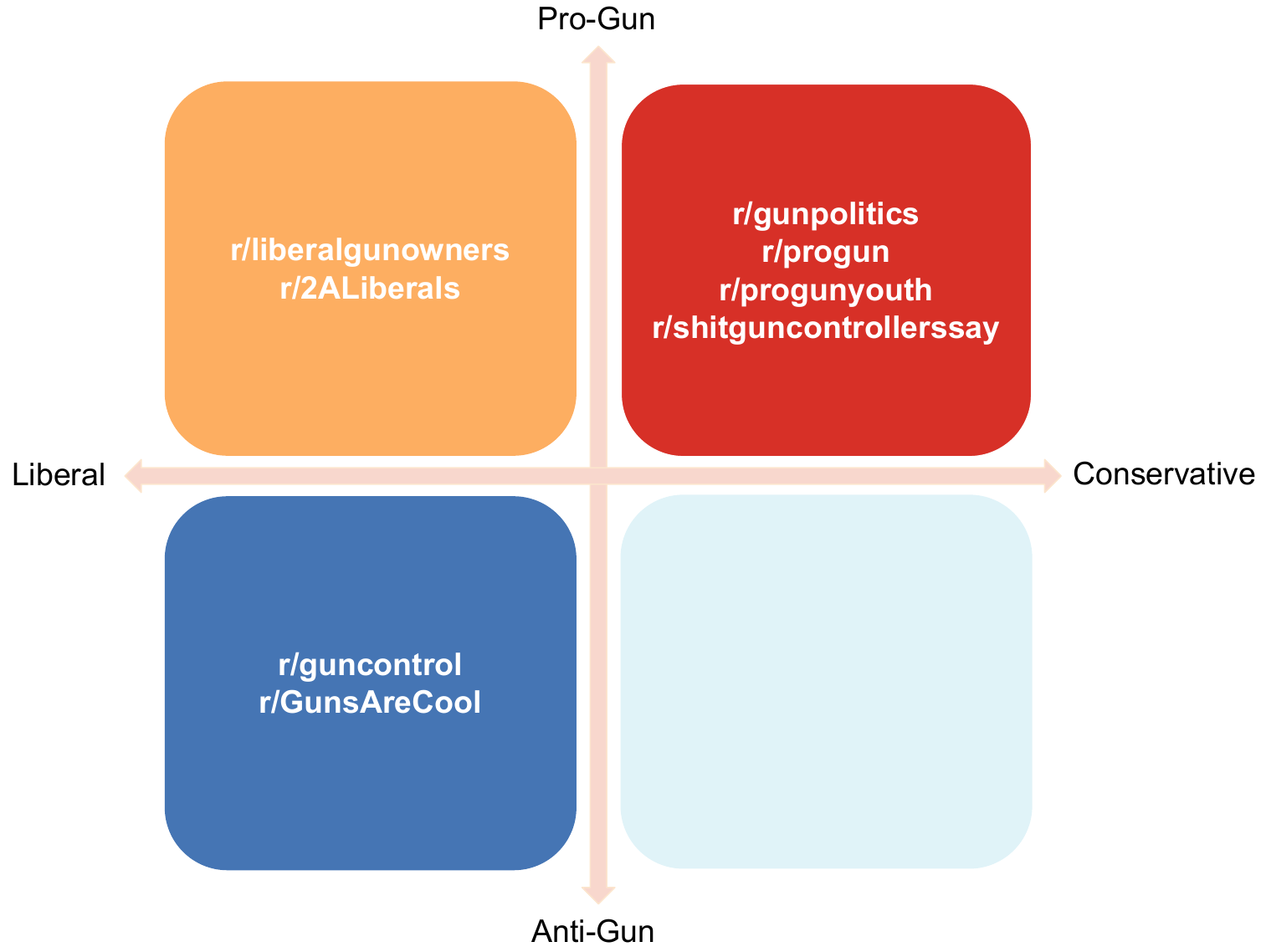}
  \caption{Subreddits represented across the axes of political ideology and advocacy toward gun control policies.}  \label{fig:gun_subreddit_alignment}
  \Description{Subreddit Alignment.}
\end{figure}

In the next phase, we categorized the eight selected subreddits on the basis of their advocacy toward gun control policies. We considered the following exhaustive labels: \textit{pro-gun} (favoring lower gun control restrictions), \textit{anti-gun} (favoring  expanded gun control restrictions) and \textit{neutral}. In addition to using cues clear from the names of the subreddits (e.g., \textit{r/progun}), authors one and four inspected the subreddit descriptions, rules and recent posting histories. Our examination revealed six pro-gun, two anti-gun and zero neutral subreddits. The inspection also pointed to recognizable political leanings apparent from context. Consequently, the authors further categorized the subreddits by political ideology. Four subreddits were found to represent conservative positions while four others corresponded to liberal leanings. No other ideological variation was found within these eight subreddits. When combined with the advocacy codes, we observe the following three categories (visualized in Figure \ref{fig:gun_subreddit_alignment}):
Conservative Pro-Gun, Liberal Anti-Gun, and Liberal Pro-Gun. 

\begin{table}[ht]
   \caption{List of subreddit categories.}
   \label{tab:lst_of_subreddit_categories}
   \begin{tabularx}{\textwidth}{lXcc}
     \toprule
     Subreddit Category & List of Subreddits & \# Posts & \makecell{Avg. Title Length\\(in Words)}\\
     \midrule
     Conservative Pro-Gun & 
      \begin{minipage}[t]{5.5cm}
        \textit{r/gunpolitics}, 
        \textit{r/progun}, 
        \textit{r/progunyouth},\\
        \textit{r/shitguncontrollerssay}
     \end{minipage} & 1034 & 13.07\\
     Liberal Anti-Gun & \textit{r/guncontrol}, \textit{r/GunsAreCool} & 600 & 12.92\\
     Liberal Pro-Gun & \textit{r/liberalgunowners}, \textit{r/2Aliberals} & 600 & 11.39\\
   \bottomrule
 \end{tabularx}
\end{table}

\subsubsection{Data}

Using the Reddit API, we extracted a set of post titles from each subreddit in Table \ref{tab:lst_of_subreddit_categories}. In total, our dataset contains 1034 samples from the conservative pro-gun category along with 600 samples from each of the liberal pro-gun and liberal anti-gun categories. Each subreddit contributed 300 samples to the dataset, with the only exception of \textit{r/progunyouth}, a relatively small subreddit which hosted 134 posts in the timeframe under consideration. The posting dates of the samples ranged from April 2022 to November 2022. 

\subsubsection{Semantic Encoding and Clustering}
In order to identify content themes from the data, we leveraged a cluster-analysis-based approach. First, we encoded each post title  
in our dataset using \texttt{all-mpnet-base-v2} embedding representations \cite{sentence_transformers_allmpnetbasev2}. The \texttt{all-mpnet-base-v2} model is a finetuned version of the MPNet model \cite{song2020mpnet}. It is particularly suited for our use case as it is finetuned on a corpus of over 1.17 billion samples of which about 726 million are sentences from Reddit \cite{Henderson2019}.

Post titles are mandatory for making submissions on Reddit, have relatively small variations in length across subreddits, and typically provide a succinct summary of the main argument of the poster. On the other hand, post bodies and comments are longer and can allow for a lot more noise. For these reasons, we decided to use titles in isolation as inputs to the pipeline; future iterations can gain from the ingestion of more features. The \texttt{all-mpnet-base-v2} model is a sentence-level encoding model that embeds an input sentence of arbitrary length to a 768-size vector. Next, we applied the HDBSCAN \cite{mcinnes2017hdbscan} clustering model to each category of samples separately. Density-based methods such as DBSCAN \cite{ester1996density} offer some advantages over distance-based methods such as K-Means \cite{macqueen1967some}. These methods are less sensitive to noise, allow higher shape flexibility of clusters and do not need to make assumptions about the expected number of clusters in the data. However, they still require parameters in the form of (1) $\epsilon$ (maximum distance between two points to be considered in the same neighborhood) and (2) the minimum number of points required to form a cluster. HDBSCAN extends capabilities of DBSCAN by removing the need for manually tuning the $\epsilon$ parameter. Since density-based methods are generally sensitive to the dimensionality of inputs, we reduced the dimensionality of the data using UMAP embeddings \cite{mcinnes2018umap} as an intermediate layer. 
UMAP or \textit{Uniform Manifold Approximation and Projection for Dimension Reduction} is a manifold learning technique for dimensionality reduction that aims to preserve the global and local structure of data. To select optimal parameters for both UMAP and HDBSCAN, we carried out grid search hyperparameters tuning using Density Based Clustering Validation (DBCV) \cite{moulavi2014density} 
 as the performance metric.\footnote{For the hyperparameter sweep, we carry out a 2 fold grid search over a range of values for the following parameters:  `number of neighbors' (UMAP), `minimum samples per cluster' (HDBSCAN), `minimum cluster size' (HDBSCAN),  `cluster selection method' (HDBSCAN) and `distance metric' (HDBSCAN)} 
 DBCV is a clustering quality metric that works on evaluating the tightness and separation of clusters and factors in both inter-cluster and intra-cluster density. The method is particularly suited as a metric for judging the clustering quality of density based clustering techniques as it is able to account for (1) non-globular cluster shapes and (2) noise as cluster labels (commonly outputted by density-based clustering methods.) DBCV values can range between -1 to 1, with values close to -1 indicating no underlying clusters and values near 1 representing highly segregated clusters. The optimal values of parameters found through the hyperparameter tuning process (independently for each subreddit category) were chosen for final clustering.

\subsection{Interview Study}
For the second analysis, we interviewed twenty people who participate in discussions of guns and gun policy on Reddit.
The study was approved by our Institutional Review Board (IRB). Participants were compensated in the form of Amazon gift cards, each valued at 20 USD. The interview length ranged between 25 to 67 minutes with an average length of 44 minutes.

\subsubsection{Recruitment}

For our second and primary analysis, we engaged in convenience and purposeful sampling, recruiting an equal number of Reddit users who posted in favor of greater gun regulation and against it. We located posts about gun politics, selecting users who created a post or comment on a topic surrounding gun interactions such that (1) they had had a history of contributing to gun conversations in the past (as evidenced by their posting history) and (2) the content contained meaningful contributions (i.e., we avoided simple reshares of links/other posts and meta discussions)

Subjects were recruited by private Reddit message. To avoid annoying participants and in accordance with our IRB protocol, we made sure that no person saw a recruitment message more than once. 

Our subjects all were recruited based on posts they made to one of the subreddits in our quantitative analysis. Our sample is a convenience sample, because we spoke to people who were willing to speak to us. It is a purposeful sample, because we recruited an equal number of people whose views are against and for greater gun regulation. Finding people who are against gun regulation (pro-gun) was easier than finding those who favor gun regulation (anti-gun). As our recruiting went on, we invested more effort in recruiting anti-gun individuals until we had a balanced sample.

Most interviewees were interviewed within a few weeks of posting content. All interviews were conducted between May and September of 2022 by authors one, two and six. 

We continued interviews until saturation was reached. 
In qualitative interviewing, saturation occurs when each new subject recruited repeats substantially the same information as previous subjects \cite{guest2006interviews}. In this study, we reached saturation at twenty interviews.
Participant background information is described in  Table \ref{tab:participant_bios}. 
To safeguard the identities of the participants, we do not directly provide details about their specific posts. Furthermore, we use light disguise to refer to our participants \cite{bruckman2006teaching} throughout the paper. As a result, while we omit sensitive details concerning our participants, it is theoretically possible for highly determined community members to discover the identities of the participants.

\begin{table*}
  \caption{Participant Biographical Information. ``LA'', ``LP'' and ``C'' in the Participant ID column refer to ``Liberal Anti-Gun'', ``Liberal Pro-Gun'' and ``Conservative'' respectively.}
  \label{tab:participant_bios}
  \begin{tabular}{cccl}
    \toprule
    Gun Stance & Subreddit Political Leaning & Highest Level of Education & Participant ID\\
    \midrule
    Pro-Gun & Liberal & High School & LP1 \\
    Pro-Gun & Conservative & Graduate & C1\\
    Pro-Gun & Conservative & Graduate Degree (Incomplete)& C2\\
    Pro-Gun & Conservative & High School & C3 \\
    Pro-Gun & Conservative & High School & C4\\
    Pro-Gun & Liberal & Bachelor's & LP2\\
    Anti-Gun & Liberal & High School & LA1 \\
    Anti-Gun & Liberal & Graduate & LA2\\
    Anti-Gun & Liberal & Bachelor's & LA3\\
    Anti-Gun & Liberal & Declined to Answer & LA4 \\
    Anti-Gun & Liberal & High School & LA5 \\
    Anti-Gun & Liberal & Graduate & LA6\\
    Pro-Gun & Liberal & Graduate & LP2\\
    Pro-Gun & Liberal & Bachelor's & LP3 \\
    Pro-Gun & Conservative & High School & C5\\
    Anti-Gun & Liberal & Graduate & LA7\\
    Anti-Gun & Liberal & Graduate & LA8 \\
    Anti-Gun & Liberal & Bachelor's & LA9\\
    Anti-Gun & Liberal & Associate's & LA10\\
    Pro-Gun & Conservative & Bachelor's & C6 \\
    \bottomrule
  \end{tabular}
\end{table*}

\subsubsection{Interviews}
Next, we conducted semi-structured interviews with users to learn about their experiences. The semi-structured design allowed for maximum flexibility in delving into discussions surrounding motivations and decision-making.  During each interview, we started with questions concerning the interviewee's general experience being on the subreddit and then followed up with queries on the specific interaction at hand. Discussions focused on the affordances of Reddit as a platform for political discussion, including the kinds of features that encourage users to contribute, but also any features, if any, that prevent cross-ideology discourse. The conversations provided several cues for possible changes to improve the quality and frequency of political engagements online. In addition, questions about individual political leanings and demographic background were asked at the end of the interview.

\subsubsection{Data Analysis}
Once the interviews were collected, the first author conducted qualitative coding of the interview data. An iterative approach was followed to consolidate and recode the data based on observable recurring patterns. In the first round of coding, the first author generated a rough set of over 100 codes. The codes were gradually distilled to a set of 11 final codes, which were discussed by the research team. Thematic analysis was carried out to organize the coded data into themes \cite{braun2006using}. The resulting themes were used to inform our findings.

\subsubsection{Positionality Statement}
All authors generally hold liberal views and support an increase in regulation of guns. Two authors are United States citizens, two are Indian nationals and two are from the People's Republic of China. Given that the study focuses specifically on gun politics in the United States, we recognize that nationality can influence perspectives on the topic. No author contributed to any of the gun politics subreddits discussed in this paper prior to the study. Therefore, the experiences and positions of the authors are different from many who participate in these communities. A majority of the authors felt their views on gun regulation grew more nuanced after completing the study.

\section{Results}

\subsection{Content Analysis: Title Cluster Themes}
\begin{figure}[h]
\centering
\begin{subfigure}[b]{0.45\textwidth}
    \centering
    \includegraphics[width=\textwidth]{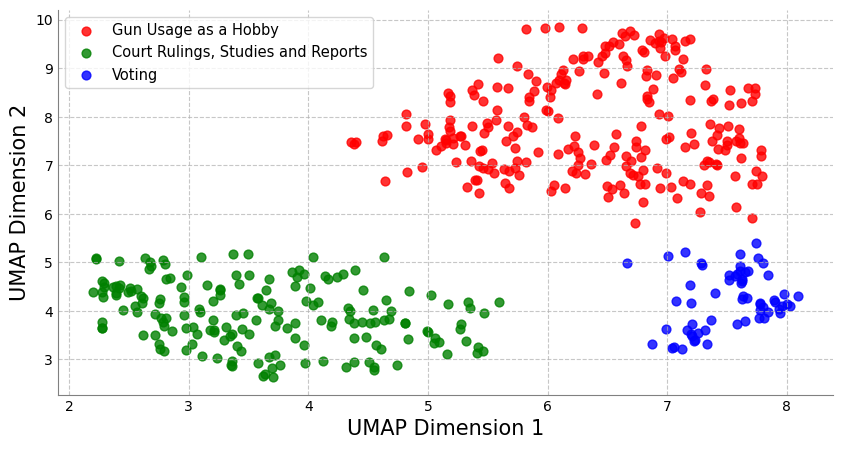}
    \caption{Liberal Pro-Gun Clusters}
    \label{fig:lib_pro_gun_clusters}
\end{subfigure}
\hfill
\begin{subfigure}[b]{0.5\textwidth}
    \centering
    \includegraphics[width=\textwidth]{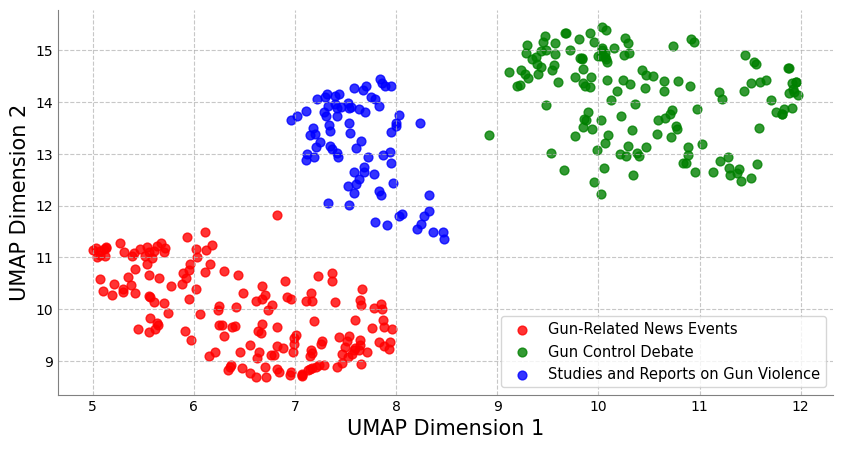}
    \caption{Liberal Anti-Gun Clusters}
    \label{fig:lib_anti_gun_clusters}
\end{subfigure}
\newline
\null\hfill
\begin{subfigure}[b]{0.5\textwidth}
    \centering
    \includegraphics[width=\textwidth]{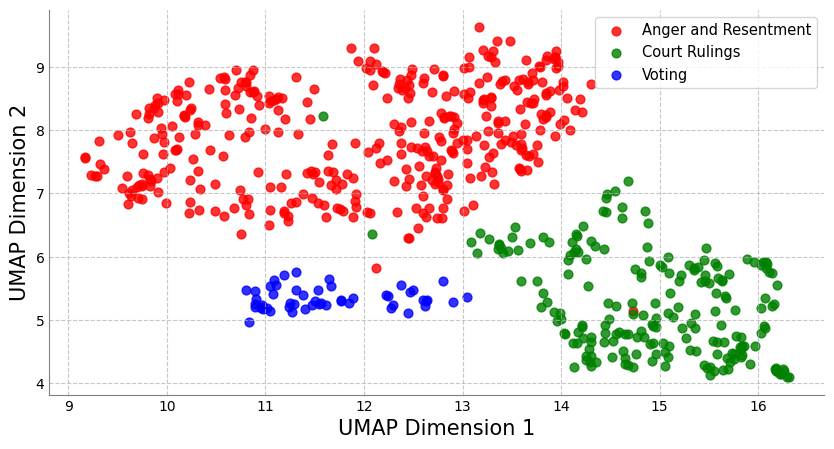}
    \caption{Conservative Pro-Gun Clusters}
    \label{fig:cons_pro_gun_clusters}
\end{subfigure}
\hfill\null
\caption{Major title clusters identified across Liberal Pro-Gun, Liberal Anti-Gun and Conservative Pro-Gun subreddits. The dimensionalities of UMAP embeddings have been reduced to 2 dimensions create the visualizations.}
\label{fig:subreddit_cluster_viz}
\Description{Major title clusters.}
\end{figure}

We discuss interpretations of the cluster analysis for each subreddit category. 
To control for noise, we interpreted only those clusters that have a prevalence (contribution of the cluster to the sample set in percentage) higher than 5\%. The clustering originally yielded 6, 3 and 3 clusters for the categories of ``Conservative Pro-Gun'', ``Liberal Pro-Gun'' and ``Liberal Anti-Gun'' respectively. The aforementioned prevalence filter segregated out 3 small-sized clusters from Conservative Pro Gun to yield 3 major clusters for each category.
While there is some amount of intra-cluster variance, we manually coded the clusters based on observable recurring themes.
To validate the quality of clusterings we extracted the UMAP embeddings for each title in the extracted clusters (using a consistent set of hyperparameters\footnote{To ensure consistency across the three clusterings, we used a common set of hyperparameters. These included default values of 2 for 'number of components' and 15 for 'number of neighbors,' along with cosine distance as the distance metric.}) and computed the final DBCV values for each category. These were found to be 0.26, 0.78 and 0.67 for Conservative Pro-Gun, Liberal Pro-Gun and Liberal Anti-Gun respectively, reflecting reasonable overall cluster quality for each category (DBCV values can range from -1 to 1, with higher values indicating better clusterability). In addition, we generated 2-dimensional projections of the clusters to visually inspect the results (Figure \ref{fig:subreddit_cluster_viz}) which provided a second layer of verification.

\begin{figure}[h]
  \centering
  \includegraphics[width=0.8\linewidth]{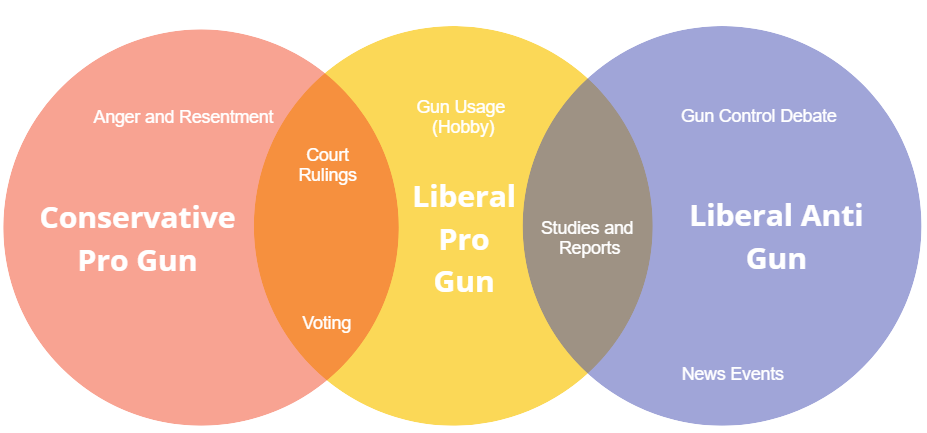}
  \caption{Symbolic representation of title cluster themes. Liberal Pro-Gun clusters intersect partially with both Conservative Pro-Gun and Liberal Anti-Gun clusters.}  
  \label{fig:quant_cluster_themes}
  \Description{Symbolic representation of title cluster themes}
\end{figure}

The clusters are also symbolically visualized in Figure \ref{fig:quant_cluster_themes}. The figure indicates the three categories (``Conservative Pro-Gun,'' ``Liberal Pro-Gun,'' and ``Liberal Anti-Gun'') along with the extracted cluster themes. Overlapping themes have been marked within the intersections of the cluster representations. For instance, \textit{voting} was noted to be a common theme between Conservative Pro-Gun and Liberal Pro-Gun. The Conservative Pro-Gun cluster \textit{Court Rulings} and the Liberal Anti-Gun cluster \textit{Studies and Reports} were found to be a combined cluster represented in Liberal Pro-Gun as \textit{Court Rulings, Studies and Reports}. They have been indicated separately in Figure \ref{fig:quant_cluster_themes} to simplify visualization. The clusters, along with representative examples and respective prevalence (a ratio of the number of posts covered by the cluster to the total size of the category) values are described in Table \ref{tab:content_clusters}.

\begin{table}[ht]
   \caption{Content cluster themes for each subreddit category. 'Prevalence' represents the ratio of the number of posts covered by a cluster to
the total count of posts in the category to which the cluster belongs.}
   \label{tab:content_clusters}
\begin{tabularx}{\textwidth}{p{1.67cm}Xp{5.3cm}c}
\toprule
Category & Cluster Theme & Examples & Prevalence \\ \midrule

Conservative Pro-Gun
    & Anger and Resentment
        & \textit{``The dishonesty makes me seethe.''}, \textit{``Get a load of this BS.....''}, \textit{``Here is another super post of idiots for yall to enjoy''} & 37.2\% \\ \cmidrule{2-4} 
    & Court Rulings
        & \textit{``The Supreme Court's ruling in Bruen was a big win for gun rights''} & 20.0\% \\ \cmidrule{2-4} 
    & Voting
        & \textit{``Voted from a 2A Sanctuary.''}, \textit{``How is PA looking for general elections?''} & 5.0\% \\
        
\midrule

Liberal Anti-Gun
    & Gun-Related News Events
        & \textit{``Hunter wipes out bighorn sheep for fun (Mesa Co., CO)''} & 28.3\% \\ \cmidrule{2-4}
    & Gun Control Debate
        & \textit{``What mistakes are being made about gun control in this twitter rant?''} & 24.2\% \\ \cmidrule{2-4} 
    & Studies and Reports on Gun Violence
        & \textit{``Red States have a serious homicide problem: Where there are more guns, there is more homicide.''} & 13.7\% \\

\midrule

\makecell[l]{Liberal \\ Pro-Gun}
    & Gun Usage as a Hobby
        & \textit{``My Sig P226 Mk25. Upgrades in Comments'' } & 41.6\% \\ \cmidrule{2-4}
    & Court Rulings, Studies and Reports
        & \textit{``Guns on the Streets: Thousands of stolen guns in city help fuel violent crime''} & 30.9\% \\ \cmidrule{2-4} 
    & Voting
        & \textit{``I’ve Become A Single Issue Voter''} & 10.1\%     \\

\bottomrule
\end{tabularx}
\end{table}

\subsubsection{Conservative Pro-Gun}
Conservative pro-gun posts revolve around three major themes---\textit{Anger and Resentment},
\textit{Court Rulings}  and \textit{Voting}.
The first cluster consists of several examples of title posts that reflect anger or resentment toward an individual or group, often referencing another post on a different subreddit (\textit{``Here is another super post of idiots for yall to enjoy''}, \textit{``The dishonesty makes me seethe.''})   

The second cluster focuses on court rulings and laws, pertaining to legal aspects of gun ownership (\textit{``Supreme Court orders lower court to reconsider Massachusetts gun control law''}). Finally, the last cluster comprises gun politics through the prism of voting, government, and mobilization.  Titles in this cluster pertain to discussions around elections and the voting process, especially relating to issues involving access to guns and gun rights (\textit{``Voted for God, Country, Freedom and 2A.''}, \textit{``What this election means for the future''}).

\subsubsection{Liberal Pro-Gun}
 Liberal pro-gun posts reflect shared attributes with both conservative pro-gun and liberal anti-gun communities.   The cluster analysis revealed three themes --- \textit{Voting and Mobilization},  \textit{Studies and Reports on Gun Usage} and \textit{Gun Usage as a Hobby}. Liberal pro-gun clusters share some intersections with both conservative pro-gun and liberal anti-gun clusters. ``Court Rulings, Studies and Reports'' form one cohesive cluster within liberal pro-gun themes, but are present separately as ``Court Rulings'' for conservative pro-gun and ``Studies and Reports'' for liberal anti-gun. 

 Similar to participants in conservative pro-gun groups, members of liberal pro-gun groups also urge members to vote on issues (\textit{``All your vote are belong to Ohio''}, \textit{``Jumping on the bandwagon, go vote everyone!''}). The posts may also refer to declarations of votes themselves. (\textit{``Obligatory I voted photo from Michigan.''})

 The next cluster pertains to academic studies on gun culture. The studies raised for discussion are typically neutral or positive in how they reflect upon gun ownership (\textit{``Guns on the Streets: Thousands of stolen guns in city help fuel violent crime''}, \textit{``Study shows majority of gun owners support specific gun safety policies''})
 
 Finally, liberal pro-gun participants often discuss gun ownership as a hobby. Posts contain references to specific firearms (\textit{``My Sig P226 Mk25. Upgrades in Comments''}), requests for advice/recommendations ( \textit{``Seeking recommendations for my first handgun''}) and general discussion points around hobby usage of guns ( \textit{``Does anyone else find rifles boring?''}). For conservative pro-gun groups, hobby and policy discussions are typically in separate groups (and we chose to study policy groups). The smaller liberal groups typically merge both policy and hobby conversations in each group.
 
\subsubsection{Liberal Anti-Gun}
Liberal anti-gun posts deal with the themes of --- \textit{Studies and Reports on Gun Violence}, \textit{Gun-Related News Events},  and the \textit{Gun Control Debate}. Like liberal pro-gun groups, liberal anti-gun groups are interested in academic studies, reports, and statistics around gun violence. With few exceptions, the implications of these study descriptions align with the political stance of the anti-gun community (e.g., \textit{``Red States have a serious homicide problem: Where there are more guns, there is more homicide''}). Discussion by liberal users values evidence/research based discussion on the topic of gun politics, though the specific studies chosen tend to reflect the group's values.

Sharing of News events surrounding specific instances of gun violence significantly contributes to liberal anti-gun interactions. Such posts re-mobilize the community toward a central community goal of strengthening gun control measures in the country. Many posts pertain to major events such as mass shootings, but others also bring up smaller-scale events ranging from accidental firings to home shootings. Commonly, the framing of the posts highlight the aspect of ownership: shooters are clearly labeled as gun owners when applicable, as opposed to more general references such as ``man,'' ``woman,'' or ``person'' (\textit{``Gunowner walks into a casino, shoots a man twice in the head, walks out and shoots himself (Anaconda, MT)''}). This is consistent with prior work \cite{demszky2019analyzing} that has pointed to differences in framing in discussions of shooting events across ideological divides. Posts, particularly in instances where the incident was non-lethal, are at times written in a mocking tone as an indirect attack on gun proponents (\textit{``Naked responsible gun owner shows up to family barbecue, shoots someone''})

The last cluster pertains to debates, questions and discussions specific to gun control.
Often, these posts are questions posed to the community (\textit{``What are some really good Anti-gun/pro-gun-control arguments have you heard?(Sources needed please)''}). 

\subsection{Interviewee Background and Interest in Guns}
In the following sections, we examine the results of our qualitative investigation, which give insights about underlying experiences, motivations, and challenges of users that shape the patterns of discussion that emerged in our quantitative analysis.

Differences in backgrounds and exposure to gun politics shape people's core views and influence the types of discourse they engage in.

\subsubsection{Introduction to Guns}

Many participants first interacted with gun culture at a young age. In particular, several pro-gun participants were introduced to guns by family members (especially fathers) as children. These came about as a result of sport and family hunting-related trips. 

\begin{quote}
    \textit{I got a BB gun when I was seven years old. I think I got a hunting permit in my first shot gun that I used with my father when I was 14 and had to go through firearms training. (C2)}  
\end{quote}

Subjects said this early exposure to guns sparked an interest in gun-related discourse, and shaped their positions on gun issues.

\begin{quote}
    \textit{Starting at the age of 12, we had a weekly skeet shooting competition between, you know, me and my dad and my brothers and some of his friends from his work. So they've always just been kind of part of my life. And I've always felt that when you hit 18, you should be able to buy your own gun. (C4)}  
\end{quote}

Exposure to cultural influences formed another gateway to developing an interest in guns. Participants mentioned watching movies and playing video games as a driving force toward learning more about guns (C3, C5). Other reasons invloved an interest in engineering/mechanical aspects along with the historical context of gun usage (LP1, LP3, C3). 

\subsubsection{Initial Involvement in Gun Politics}

From a political perspective, several factors fostered deeper engagements with gun politics. These ranged from family dinner-table conversations (LA8, LA10) to assessing the impact of gun culture differences across state, country or rural/urban dividing lines (LA1, LA2, LP2) to being exposed to political content online (C3, LA9).

For multiple liberal anti-gun subjects, interest in gun politics came about after learning about one or more mass shooting events (LA3, LA5, LA6). Frequently, for such participants the emotional impact of the event were a driving force in their wanting to engage with like-minded users.

\begin{quote}
    \textit{My kid was about eight, or nine when Newtown happened. And I was like, wow, this is this would just ruin me if that happened to me. As a parent, can you imagine? Right?   That was the first time I showed up to read. I was looking around, I guess. And I found Reddit, the gun politics subreddit. (LA3)}  
\end{quote}

In some cases, gun violence related events in the direct surroundings of the participants influenced perspectives (LA1, LA7). One participant spoke about a personal tragedy that had a lasting impact on their psyche and consequently their desire to engage actively in gun discussions.

\begin{quote}
    \textit{In the early 2000s, my uncle who had addiction problems, had a confrontation with the police. And he was armed. He refused to relinquish his firearm, and the police shot him and he died. After that, I also became very interested in just changing things so that that kind of situation doesn't happen.  (LA2)}  
\end{quote}

\subsubsection{Evolution of Interest in Guns}

For several subjects, stances toward guns have remained unchanged over the course of their lives. Some participants, however, stated that their positions altered over time. The catalyst to these changes often was exposure to statistics or other quantitative metrics that the subjects came across with regard to gun politics, particularly on measures that related to gun violence. For instance, LA8 cited an online post by Michael Shermer (publisher of the \textit{Skeptic} magazine \cite{Skeptic}) as a major contributing factor toward them modifying their stance.

\begin{quote}
\textit{When I was younger, I was very much on the other side of this issue. And I've slowly come around to being more aligned on the gun control side... It was a post on ... Facebook or Twitter... about how much gun violence there is...  that really resonated with me and actually made me change.  (LA8)}
\end{quote}

In some cases, positions changed partially, with participants relaxing the degree to which they held their views. For instance, LP2 discussed how they softened their stance on gun ownership due to increasing cases of gun violence. 

\begin{quote}
    \textit{Yes, I used to be very much more of a hardline, anti regulatory person. But with the ongoing problem of gun violence, it's pretty apparent that something needs to be done about it. So my focus on this issue is more thinking about ways we can preserve gun ownership while addressing the problem. (LP2)}  
\end{quote}

\subsection{Engaging with Opposing Views}
Despite the segregation of views across political lines between subreddits, a large number of participants stated that they are open to interactions with those holding opposing views (LP1, LP2, LP3, LA1, LA2, LA3, LA9, LA10, C1, C2, C3, C4, C5). While issues around social desirability bias \cite{nederhof1985methods} in interview responses cannot be fully discounted (see Section \ref{limitations} on limitations), the findings point to the potential for designing new spaces for cross-ideology dialogue. Notably, to a subset of our interviewees, cross-ideology interactions provided them opportunities to either expand their understanding of gun politics or to simply hold more meaningful conversations around the topic.

 \begin{quote}
     \textit{Online, I'm usually only engaging with people that I strongly disagree with. Because it's just really pointless to just virtue signal with your friends in the echo chamber. (LA10)}  
 \end{quote}

 To others, these interactions also served as a mechanism to influence the opposing side toward changing their views.

 \begin{quote}
     \textit{I usually just seek out debates with people I disagree with... because I find it more interesting to challenge my own perspectives, as well as, potentially, the side benefit could be that I could convince them to my position. (LA9)}  
 \end{quote}

\begin{quote}
     \textit{I was hoping I could change a few minds or at least plant a seed. (LP2)}  
 \end{quote}

\begin{quote}
     \textit{I don't think that you can convince people to agree with you, if you don't listen to what they have to say and understand why they disagree. So that's generally the way that I further familiarize myself with anti gun perspectives, it's by talking to them so that they stop trying to vote away our rights. (C3)}  
 \end{quote}

However, participants also discussed impediments to having these interactions. Typically, these were presented in the form of apprehensions about conversations turning aggressive to the point of being hostile. 
Others harbored concerns about conversation partners being disingenuous, making any prospective interaction futile.

\begin{quote}
     \textit{If I'm going into a group of people, a bunch of people that disagree with me, it's best to make sure that I know people are civil first. (C5)}  
 \end{quote}

\begin{quote}
     \textit{Even though we disagreed on a lot of topics, it was very constructive, very positive. It was a great experience, but I had to focus on just the conversation between me and those two other people, because the rest of it was generally just pretty toxic. (C4)}  
 \end{quote}

\begin{quote}
     \textit{I would prefer to have a discussion with people who have diverse views but only if everyone will approach the subject with an open mind and be sort of intellectually honest about things. (LA2)}  
 \end{quote}

Participants expressed that the fear of conversations becoming uncivil persisted in offline settings as well, particularly in scenarios that involved members of their close circles.

\begin{quote}
     \textit{It can be a very civil conversation, even if we strongly disagree... I have other family members where I know that it likely will not be a very civil conversation, because we disagree. And so lately, I've just been avoiding the politics conversation, although I definitely have had those conversations in the past. (LA8)}  
 \end{quote}

\begin{quote}
     \textit{Yes, I'd really like to [talk to to them], especially the friends I disagree with, but I also don't want to step on any toes. (LA1)}  
 \end{quote}
 
That said, in comparison to offline discussions on controversial topics, some distinctive features of online interactions (particularly on Reddit) can exacerbate the likelihood of discussions turning toxic. Some specific features that were reported were anonymity, latency and text-only communication modalities. For instance, C4 referred to the issue of users feeling more comfortable with attacking individuals, since it is difficult for them to face direct harm due to anonymity and a digital landscape. 

\begin{quote}
     \textit{If you're really inflammatory with somebody in person and trying to push all their buttons, you're gonna get punched in the face. Online you don't have that reservation, right. You can be as as hateful and caustic as you want, and just walk away from it. (C4)}  
 \end{quote}

LA2, on the other hand, discussed how access to audio-visual modalities such as voice and appearance along with a lower response time in real world settings made communication more engaging. 

\begin{quote}
     \textit{I think when you look at people and actually interact with them, it's much more humanizing. It's not just words on a screen. It's a real person. And generally speaking, in my experience, actually having a discussion with someone, where you can have immediate feedback and see and hear them, tends to make people more engaged and less hostile. (LA2)}  
 \end{quote}

In summary, participants generally  expressed openness toward interacting with those they disagreed with on gun politics. However, expectations of conversations turning toxic deterred participants from fully engaging. From an online perspective, specific features of platform discourse were reported that potentially amplified these issues.  

\subsection{Engaging with Like-Minded Subreddits} \label{like-minded}

Although many interviewees did express interest in engaging in cross-ideology discussions, some interviewees discussed reasons why they typically limited their participation to like-minded subreddits (C3, C5, LP1, LP2, LP4, LA2, LA3, LA6, LA7, LA10). Our interview findings suggest a two-pronged effect where these users are being pushed away from opposing subreddits while feeling compelled to seek out like-minded subreddits.

Interviewees generally cited negative characteristics of opposing subreddits as reasons for why they avoid interacting with them. These characterizations primarily referred to pro-gun spaces on Reddit. LA2 described pro-gun subreddits as places where anti-gun views are \textit{``seen as traitorous, un-American, a direct threat to those individuals''} and \textit{``[members] react so negatively to any other opinions that it's not worth trying to have the discussion.''} LA4 described how the prevalence of brigading (see Section \ref{downvotes}), or mass downvoting, deterred their participation in pro-gun subreddits. These sentiments were echoed even by liberal pro-gun participants, who noted that these traits were associated with the conservative pro-gun crowd specifically. For example, LP2 describes trying to get away from \textit{``all of the craziness that's associated with… extremist right-wing ideology that you get in regular gun circles [on Reddit].''} LP4 mentioned that \textit{``there really is an element of paranoia and aggression among pro Second Amendment people on Reddit that I don't necessarily think is healthy.''} On the other hand, conservative pro-gun interviewees expressed how they did not engage with outgroup communities out of a desire to avoid conflict or debate. C3 describes how engaging with anti-gun spaces as a pro-gun individual is a hostile and demanding endeavor:

\begin{quote}
    \textit{It’s a very different type of dialogue, where you're always on the defensive, and you're always needing to give an explanation. And you never really have other people to back you up. So it sort of forces you to know what you're talking about, and to be more well informed. (C3)}
\end{quote}

Similarly, C5 spoke about how they actively avoided subreddits where there were more cross-cutting discussions:

\begin{quote}
    \textit{I'm a very competitive person. And if I get into a debate on politics, it'll go on for a long time… topics there can get very heated and if I do get into an argument like that, unfortunately it's all I think about for quite a while and it's not quite healthy for me to do that. (C5)}
\end{quote}

On the other hand, interviewees described looking for a sense of belonging among people with common experiences when talking about why they tended to gravitate toward like-minded subreddits. This was true for users across the political and ideological spectrum. LA2 described how they \textit{``like knowing that I'm not alone''} in feeling frustrated about the current situation regarding guns in the United States. LP1 mentioned how they felt that their perspective as an LGTBQ+ gun owner would be more readily accepted on \textit{r/liberalgunowners} because of the subreddit’s \textit{``underlying moral foundation''} and its \textit{``introspective''} nature:

\begin{quote}
    
    \textit{I think that most people on [r/liberalgunowners] would agree with me... one thing that I see there very often is armed minorities harder to oppress. Armed queers don't get bashed. Black gun ownership is a human right. (LP1)}
\end{quote}

C3 described how they specifically sought out \textit{r/progunyouth} as a place of belonging as a young person with a pro-gun perspective:

\begin{quote}
    
    \textit{Throughout high school, it was basically just me and a handful of other people who are openly pro-gun. So there are definitely like assumptions that people make about you... So I was interested in a subreddit where other people had shared that experience. (C3)}
\end{quote}

Interviewees described how the hostile or otherwise disagreeable environments of outgroup subreddits drove them away while feeling the need to seek belonging in subreddits of people who shared like-minded views on gun politics. Although this two-pronged phenomenon may indeed contribute to more polarized spaces, most interviewees who were influenced by these push-pull effects nevertheless expressed interest in engaging in cross-partisan discourse on gun politics.

\subsection{Post Patterns}
As a part of the recruitment process, participants were contacted on the basis of their posting history. Thus, during the interviews, we asked participants to tell us about any of their recently posted content related to gun politics. This was followed primarily by questions along the lines of: \textit{What motivated you to post this?} and \textit{How do you feel the post was received?} Often, multiple posts in any given interview were discussed in this format along with follow-up questions. Learning from this section helps us answer \textbf{RQ2}.

\subsubsection{Purpose of Post}
Participants reported a mixture of motivations behind posting content. Some declared that posting content was largely a means to share ideas (LP1, LA3, LA5, LA7, LP4) --- either to provide the community with information or add their own perspectives to conversation, while not explicitly seeking to change people's minds.

\begin{quote}
    \textit{I don't think it was trying to influence anybody...  I think it was just expressing my own perspective on the matter. (LP1)}  
\end{quote}

\begin{quote}
    \textit{[I wanted to provide] information, just inform... I'm a classical liberal, Democrat, and I think an informed populace makes better voting decisions and keeping the people informed is important. (LA3)}  
\end{quote}

A different subset of participants were motivated by a need to influence those with differing opinions and persuade them to consider modifying their stance (LA1, LA2, LA8, LA10). It is worth noting that all participants with this position leaned toward an anti gun stance. 

\begin{quote}
    \textit{I was probably trying to persuade. I suspect there are people in the gun control subreddit who would want to get rid of all guns everywhere. And I think that might be a noble ideal, but it just seems very unrealistic to me. And so urging for safety and training, I think, is the probably the step I'd want to promote.  (LA8)}  
\end{quote}

Additionally, a group of participants felt a need to post in order to emotionally express themselves or vent to the community (C3, C6, LA6). The desire was fueled by a general sense of frustration or a recent tragic event (such as a mass shooting). For example, C3 recalled ``venting'' to Reddit about their frustration with the popular narratives surrounding the 2022 school shooting in Uvalde, Texas:

\begin{quote}
    \textit{So this was either a day or maybe a few days after the school shooting in Texas recently... It's the most horrific news that we've gotten in the last decade. And what I think made it particularly disturbing was how hours after this reached the news, people were politicizing it and using it to push their perspectives. And that really didn't sit well with me. So I basically posted about how a lot of people assume that if you don't support gun control, you don't care about the victims of mass shootings, which is a very upsetting assumption for people to make.  (C3)}  
\end{quote}

\subsubsection{Post Reception}
Participants experienced wide ranging degrees of acceptance by the respective communities. They judged acceptance based on the nature of textual feedback (in the form of comments) along with the votes scores they received on their posts. Community reactions appeared to be affected by the timing of the post (e.g., if there had been a recent shooting event) and the degree of alignment between the perceived ideological leaning of the poster with that of the community at large, in addition to the post content itself.

Unsurprisingly, several participants (LA1, LA2, LA4, C1, C2, LP1, LP4) reported a positive reception by the community, in cases where their own political stance matched with the community. For instance, LA1 described how \textit{``there were a lot of comments that were in agreement... giving out their own examples.''}

In some scenarios, reception was mixed or even negative. For instance, LA5 discussed the role of brigading causing their post to be heavily downvoted. Here, the participant presented a liberal anti gun view on an aligned (largely liberal anti gun subreddit) which was brigaded. 

\begin{quote}
    \textit{Obviously, not [received] very well, because it's been getting mass downvoted. So one of the admins did a comment on it, where we're saying that the subreddit's been gaining a lot of bot downvotes, and they've done nothing about it for about a year.  (LA5)}  
\end{quote}

\subsection{Reddit Gun Politics Communication: Impact of Interaction Features}
In this section, we discuss our findings on how various features of Reddit's interface affect cross-ideological communication. The interviews indicate that several design elements encourage formation of echo chambers.

\subsubsection{Downvotes and Brigading} \label{downvotes}

Reddit supports a binary upvote/downvote content rating system. Users leverage ratings to align conversations along the community’s norms and goals. Upvotes help draw attention toward specific posts or comments since such content is bubbled to the top of the page. On the other hand, downvotes assist with cleaning contributions that might fall under categories such as spam or hate speech. As such, downvoting can also be viewed as a form of soft-reporting, since downvoted content can be filtered easily for further review by moderators. Finally, posts that are downvoted more than 5 times (score of -5)  are “collapsed” in the thread, hindering easy visibility of the content.

Several subjects (LP1, C3, C4, LA1, LA3,  LA4, LA5, LP2, LA9, C6) brought up the Reddit  downvote feature as one that hinders cross-ideology communication. Normally, the act of downvoting can come in the form of lower scores given to views conflicting with norms. 

A form of mass downvoting called  \textit{brigading} was frequently cited in the context of inter-community conflicts as a strategy to attack perceived opponents. In this scenario, content that aligns with community norms is downvoted by outsiders to dissuade discussion.
    
\begin{quote}
\textit{I don't participate on any others [subreddits] because they are particularly [affected by]... brigading... where we have an online army just downvoting your posts... just harassing people who are for gun control. (LA4)}
    
\end{quote}

\begin{quote}
\textit{Unfortunately, like I mentioned we're often brigaded by pro gun activists who will strawman us, they will use whataboutism.\footnote{A fallacy that ``attempts to defend against criticism by turning a critique back
at the accuser'' \cite{dykstra2020rhetoric}} Anything except arguing the actual facts. (LA4)}
\end{quote}

Other participants reported the longer term and wider-scale effects downvotes had on engagement. In particular, the feature inhibits platform-wide participation, since ratings affect the total karma of a user. In effect, posts that are misaligned with community norms are likely to receive downvotes, which in turn adversely impacts karma of non-aligned users, affecting their reputation and their ability to effectively communicate across other subreddits. Furthermore, prior work suggests that users disregard community guidelines for voting/downvoting \cite{graham2021sociomateriality}, compounding the extent of the problem. Participant C3 expressed how this phenomena related to the polarization of Reddit communities.

\begin{quote}
\textit{The longer the discussion lasts, the more karma that the person who disagree loses, and they don't actually persuade anybody... And it just means you have echo chambers, where it's all pro gun people or all anti gun people, and no one understands where the other is coming from.} (C3)
    
\end{quote}

LP2 expressed the need of agreement/disagreement options distinct from upvotes/downvotes for users to express or withdraw support, differentiated from reporting offtopic, illegal or unproductive comments:

\begin{quote}

\textit{I think a system of having an ``I agree, I disagree'' with this opinion, sort of like Reddit's upvotes and downvotes is good, because it lets people feel like they can agree with an opinion without needing to post their own stuff... But on the other hand, what Reddit does is, if you have too many down votes, they literally collapse the thread. And I don't think that's very good for encouraging dissenting voices and having productive conversation. (LP2)}
    
\end{quote}

\subsubsection{Anonymity}
User accounts on Reddit are pseudonymous. While facets of identity can be inferred due to context collapse \cite{triggs2021context}, usernames or other public account details do not need to be linked to real-world identity. Several participants brought up anonymity as a primary feature affecting quality of both intra-subreddit (engagement of users within subreddit communities they normally contribute to) and inter-subreddit (engagements with subreddits with opposing views).

Several remarks were made on the downsides of the feature (LP2, LA6, LP2, C2), often with regards to the general theme of anonymity lowering inhibitions to behave appropriately.   

\begin{quote}
 \textit{It allows people to be more uncivilized, with fewer consequences. And so I've seen people who had said things on Reddit that... I don't think most people would ever say to another person in real life, or if you knew who the other person was, or if they knew who you were.} (LA2)
\end{quote}

\begin{quote}
 \textit{It's keyboard warrior syndrome, where somebody feels uninhibited and protected by their anonymity of a screen and a keyboard and can just be generally nasty.} (C2)
\end{quote}

Rules around account age and karma thresholds differ across communities, leading to varied standards of membership and conversation quality. Furthermore, Reddit allows users to easily create accounts with limited checks. At the time of writing, new accounts can be created even without an email address using just a username, password and CAPTCHA on old.reddit.com. A combination of these factors enable users to establish ``sock puppet'' or ``throwaway'' accounts \cite{leavitt2015throwaway} that may be directed toward trolling or otherwise attacking communities although comprehensive experimentation needs to be conducted to provide evidence. 

\begin{quote}
 \textit{If we're talking about on Reddit, specifically, you would need to have a minimum karma threshold and a minimum account age threshold cut down on people putting in a bunch of sock accounts and going into troll people.} (LP2)
\end{quote}

With that said, some participants discussed advantages of anonymity in online communication. (LA2, LA7, LA10, LP3).  
These were particularly highlighted when participants compared Reddit against fully public platforms such as Facebook:

\begin{quote}
  \textit{Being anonymous [on Reddit] helps. You're not as subject to judgment as you are on Facebook with all of your cousins and parents and stuff. And it kind of gives you more freedom to kind of collect on issues that you care about. (LA7)} 
  \end{quote}

\begin{quote}
 \textit{I used to post more on Facebook on public forums. And the trolls on Facebook were, in my experience, were much worse and much more aggressive and much more personal than the ones on Reddit because, maybe because everyone on Reddit is anonymous, people are harder to attack.}  (LA2)  
\end{quote}

Users' perceptions of the role of pseudonymity in supporting or stifling political discourse are mixed, and further investigation is needed to understand this fully.

\subsubsection{Moderation} \label{findings-moderation}

On Reddit, users are subject to content guidelines both at the platform (termed \textit{Reddiquette} \cite{fiesler2018reddit}) and subreddit levels. As prior work illustrates, content moderation plays an important role in how a community is perceived by users. Participants shed light on strong moderation efforts as a core requirement for the enablement of cross-ideology communication (C1, C2, C3, C4, LP2, LA1, LA3, LA5, LP2, LP3, LA7, LA9, LA10). The identification and ejection of bad faith actors from conversation threads were observed as a primary goal for the moderation team. 

\begin{quote}
    \textit{I think good moderation would make it successful. Because the moderators of subreddits, for example, are volunteers, and they have their own biases. So I think if you have a person who was trained moderator, to be able to come in and encourage productive conversation, be willing to remove from conversation, people who are being really intentionally unproductive or suggesting like illegal things or something like that, just to keep everything aboveboard, that would be good.  (LP2)}  
\end{quote}

Specifically as it pertains to cross-ideology discussions, participants brought up notions of fairness while making moderation decisions to induce trust. For instance, participants stressed on the need for political impartiality to accommodate both sides of the divide. 

\begin{quote}
    \textit{I think the key is to have good, like very fair and impartial moderation.  I think with a topic that's as sensitive and as controversial as this, there's a real danger of people resorting to personal attacks  for the emotional temperature to get too high.  (LP3)}  
\end{quote}

\begin{quote}
    \textit{I think potentially more moderation of people that can't discuss things in a mature fashion would help as well. But that moderation... can't be part of the problem. It can't be heavy moderation on one side and not the other, it needs to be done fairly in respect to the way somebody is discussing a topic, not the points that they make themselves. (C4)}  
\end{quote}

Participants also desired clear rules of engagement --- codification of how and what can be said --- so that the moderation policies retain objectivity.

\begin{quote}
    \textit{A surefire way of upsetting a group of people is using inflammatory language, blanket statements and loaded statements. So, guidelines and how people say things or how to present your idea [are necessary].  (C2)}  
\end{quote}

One participant sensed that a specialized platform for cross-ideology communication might require an expanded set of responsibilities for the moderation staff. In particular, they recommended moderator-led discussion prompts to reduce the likelihood of conversations going awry right off the bat due to any perceived political biases in posts.     

\begin{quote}
    \textit{The best way to go would be to start with centralized discussion prompts. And not... have a discussion thread, but each individual post should probably be selected as and written by the moderator staff, in order to foster discussion and be in a non politically charged and neutral manner. `Cause generally... on these topics that are emotionally charged... people will post leading questions... they're trying to get people to a conclusion.  (LP2)}  
\end{quote}

\section{Discussion and Design Recommendations}
Collectively, our findings provide insights into how gun communication is carried out on Reddit, especially issues pertaining to cross-ideology dialogue. Our clustering analysis shows that users interactions on gun politics differ along topical lines, based on ideological groupings of subreddits. From the reported experiences of participants, we find that \textit{users typically avoid communicating with those they disagree with, but are open to entering into a dialogue with the opposing side}. To this end, participants reported interface features of Reddit that create impediments in cross-ideology communication. In particular, participants reported issues around rating systems, anonymity, and moderation as primary factors that influenced conversations.

Grounded in our findings, we also present design recommendations that could help enable a platform centered on facilitating civil discourse on contentious issues. Our findings seek to extend our findings from studying discussions of gun control to the wider theme of cross-ideology discussions.

\subsection{Downvotes are Problematic}
Downvotes play an important role in online discourse, providing a way for the community can directly respond to inappropriate or offensive content. However, from a political discourse standpoint, downvoted posts can trigger a sequence of negative interactions or cause a spiral of silence \cite{noelle1974spiral} that discourages legitimate conversation. For instance, prior work has shown that downvoted content attracts higher engagement while generating more negative responses from the community than upvoted posts \cite{davis2021emotional}. Downvotes can also accelerate the echo chamber effect, preventing the community from engaging with diverse opinions \cite{gaudette2021upvoting}. Currently, the ability of users to downvote content (particularly in the case of outgroup participants browsing community content) appears to strongly affect participation. Downvotes directly influence the platform-wide reputation of users by lowering their karma and therefore affecting their overall experience of contributing to the platform. 
For this reason, participants feared active participation not only in subreddits with opposing views but also sometimes in those with aligned values, where they suspected mass downvoting by outgroup users. A large-scale, coordinated version of downvoting called brigading was commonly cited as a major problem by interview participants (LA4, LA5).

Prior work has shown that downvoted content attracts higher engagement while generating more negative responses from the community than upvoted posts \cite{davis2021emotional}. Downvotes can also accelerate the echo chamber effect, preventing the community from engaging with diverse opinions \cite{gaudette2021upvoting}.
Furthermore,  prior work indicates community members do not always use community voting guidelines as intended \cite{graham2021sociomateriality}, adding another layer of complexity toward allowing for deliberative discussions. 

To address these issues, we suggest that reworking the design of the voting incentive structure might be fruitful:
\begin{itemize}
    \item A simple implementation can take the form of the removal of downvotes altogether. In this setup, posts that violate rules or do not conform to community standards can still be reported. Valid posts that might otherwise have been brigaded would stand a chance to be discussed.
    \item Alternatively, a  potential enhancement to the voting mechanism can be to clearly indicate the ratio of votes received by active in-group participants versus others.   In other words, one post might show that it is being up and down voted along ideological lines, while another might show more diverse opinions.
    Activity can be defined as some function of the number of previous contributions made and votes received in a set amount of time within the community. 
    Knowledge of the sources of votes could give participants a valuable perspective, and also be useful for moderators to assess if brigading attacks are happening within the community. 
    Reddit has already limited the number of downvotes possible on a post, to discourage mass downvoting. Hiding downvotes entirely is another option, as was done on YouTube in 2021 \cite{TechCrunchYTDislike}.
    
\end{itemize}

Making these changes to downvotes could reduce the fear of backlash associated with engaging in cross-ideology discourse and lower the barrier to participation for a broader set of users.

\subsection{Moderation}
Moderation practices can vary strongly based on political leanings and norms of a community \cite{nithyanand2017online}. Primarily, participant responses indicated that community norms discourage engagement with outgroup members. A large number of participants mentioned the importance of moderation for the smooth functioning of communities, particularly those that dealt with hot-button issues (see Section \ref{findings-moderation}). While quality moderation is important for any discussion group, it is especially important for groups focusing on controversial issues. Two design features that could help are:
\begin{itemize}
    \item Metrics of the quality of moderation (responsiveness, fairness, consistency) visible to moderators to help them improve their practices.  Specific metrics could include in-depth statistics of posts (or comments) filtered by the moderation team tagged with the nature of violations per post,  average decision response time etc.

    \item Metrics of the quality of moderation visible to users, to help them choose which communities to participate in.   Providing accessible, visual cues to users can assist them in making informed decisions about making contributions. This can also incentivize communities to improve moderation practices to attract more participants.
    
\end{itemize}

More transparency in the moderation process could encourage moderators to take a more active role in promoting civil, cross-ideology discussion, which could in turn boost user confidence in the quality of discussion within that community.

Additionally, implementing such metrics raises a host of research questions. What is ``fairness'' and how can we measure it automatically? In initial explorations, we found that large language models (LLMs) may be able to provide useful and relevant feedback to moderators and users \cite{OpenAIContentModerationGPT4Blog, li2023hot}. This is an active topic of our ongoing research.

\subsection{Special-Purpose Spaces}
The final design idea growing from our findings is that one could create special-purpose groups with the explicit goal of fostering communication across difference. Such groups could have explicit rules supporting civil discourse (C2), incorporate strong moderation to enforce those rules (see Section \ref{findings-moderation}), and try to foster social norms of civility. Self-selection of participants might also help to create a civil atmosphere, if everyone coming to the group participates because they want to engage in conversation across difference. The presence of such a group in the online gun politics discussion space could also help to reduce the push-pull effect that typically leads users to only interact with like-minded individuals, as outlined in Section \ref{like-minded}. The challenge is whether many people would be interested in joining such a group. We are exploring this in our ongoing work by creating a subreddit for civil discussion of gun issues, r/guninsights. Empirical study of the new subreddit is part of our current work.

\section{Limitations} \label{limitations}

Our participants represent a convenience sample, and are also self-selected --- we spoke to people who were willing to speak with us. We also deliberately invited interview subjects who were active participants with contributions to discussions on gun policy (not just gun hardware). Therefore, we are omitting lurkers, and people who want to talk about guns as a hobby rather than as a civic issue.  Views of lurkers and hobbyists may differ from the views of people who engage with gun policy issues.

The generalizability of our findings to other social media platforms may also be limited by platform-specific features and norms. For example, the unique thread structure of Reddit means that conversational dynamics may not necessarily mirror those of other social media sites. In addition, users of Reddit on average tend to be younger and more male, which may also impact our findings \cite{proferes2021studyingReddit}.

A third limitation is that subjects may have inferred that we are interested in cross-ideology communication. As a result, social desirability bias may play a role in our results \cite{nederhof1985methods}. This can take the form of participants reacting more positively to questions around their willingness to participate in discussions with those holding opposing views or attempting to come across as ``less extreme'' with regard to their political views. This may introduce a bias in our findings \cite{bergen2020everything}.

Our clustering analysis also has limitations. One, the dataset is constrained to a small number of subreddits, with varying amounts of activity. Second, liberal anti-gun subreddits in particular host contributions that are much sparser than other categories. As a result, our quantitative results are somewhat less reliable for those subreddits.

\section{Conclusion}

Sociologist Irving Goffman documents how we are all always performing \cite{goffman2016presentation}. When we speak before others, we have an impression we wish to make. Even in simple, every-day interactions like ordering at a restaurant, there is a complex dynamic of impression management among the parties. When we speak before others about difficult issues, the stakes are raised. People feel strongly about this issue--will I offend someone? Will someone attack me? In this light, it's remarkable that people attempt to discuss difficult issues at all.

In this paper, we have documented the process of discussing a specific controversial issue (gun politics) on a particular platform (Reddit). Using a quantitative analysis of posts from a diverse sample of discussion groups, we detail what people talk about. In interviews with members of those groups, we document how they feel about discussing controversial issues online with strangers in a pseudonymous venue.

Based on our data, we explore what specific design features of the online environment both hurt and hinder cross-ideology discourse. Possible design changes that emerge from our findings include fundamental changes to upvoting and downvoting. While feedback from peers is important, if the issue is controversial, the existing voting mechanisms can stifle diversity of thought. Might it help to separate an assessment of the quality of the content from agreement or disagreement with its substance? Could it help to provide meta-data on who approves or disapproves of content? An assessment that a post is being up/down voted along partisan lines is a different signal than a message that receives more complex reactions. Could creating special-purpose groups for bipartisan discussion lead to better discussion outcomes? As moderators struggle to help maintain civility, do they need better tools to support their effective leadership? In what ways can Natural Language Processing technology help? Would available metrics of conversation quality help moderators improve their practice? Could feedback on comment quality help participants post in a more civil fashion? There are rich possibilities in this design space that we are addressing in our ongoing work.

\section{Acknowledgements}

We are grateful to the anonymous reviewers, as well as members of the ELC and SALT labs for their feedback on the draft. This work was funded by grants from the Institute for Humane Studies (IHS) and Meta. 

\bibliographystyle{ACM-Reference-Format}
\bibliography{refv2}

\received{July 2023}
\received[revised]{January 2024}
\received[accepted]{March 2024}

\appendix

\end{document}